\documentclass [useAMS]{mn2e}
\usepackage {graphicx}

\begin{document}

\title{BZ-MC-BP Model for Jet Production from Black Hole Accretion Disc}
\author[]{Ding-Xiong Wang$^{*}$, Yong-Chun Ye, Yang Li and Zhao-Jiang Ge \\
$$ Department of Physics, Huazhong University of Science and Technology, Wuhan,430074,China \\
$^*$ Send offprint requests to: D.-X. Wang (dxwang@hust.edu.cn) }
\maketitle

\begin{abstract}

Three energy mechanisms invoking large-scale magnetic fields are
incorporated in a model to interpret jet production in black hole
(BH) systems, i.e., the Blandford-Znajek (BZ) , the magnetic
coupling (MC) and Blandford-Panye (BP) processes. These energy
mechanisms can coexist in BH accretion disc based on the magnetic
field configurations constrained by the screw instability, provided
that the BH spin and the power-law index indicating the variation of
the magnetic field at an accretion disc are greater than some
critical values. In this model the jets are driven by the BZ process
in the Poynting flux regime and by the BP process in the
hydromagnetic regime, being consistent with the spine/sheath jet
structure observed in BH sources of stellar and supermassive size.

\end{abstract}


\section{INTRODUCTION}

As is well known, jets exist in many astronomical cases, such as
active galactic nuclei, quasars and young stellar objects. Different
theoretical models have been proposed for acceleration and
collimation of jets, which can be divided into two main regimes.
Energy and angular momentum are carried by both the electromagnetic
field and the kinetic flux of matter in the hydromagnetic regime,
and those are carried predominantly by the electromagnetic field in
the Poynting flux regime (Ustyugova et al. 2000; Lovelace et al.
2002). Blandford \& Znajek (1977) proposed firstly that jets from
AGNs can be powered by a rotating black hole (BH) with a large scale
magnetic field threading its horizon. Later, Blandford \& Payne
(1982, hereafter BP82) suggested that an outflow of matter can be
driven centrifugally by large-scale magnetic fields anchored at the
disc surface. These two mechanisms are usually referred to the BZ
and BP processes, respectively. As argued in BP82, an outflow of
matter can be driven centrifugally from the disc, provided that the
angle of the poloidal magnetic field with the normal of disc surface
is greater than a critical value, i.e., $\alpha_{_{FL}}>30^{0}$. The
BZ and BP processes belong to the Poynting flux and hydromagnetic
regimes, respectively.

Recently, much attention has been paid to the magnetic coupling (MC)
of a rotating BH with its surrounding accretion disc, and this
mechanism is referred to as the MC process, which can be regarded as
a variant of the BZ process (Blandford 1999; Li 2000, 2002a; Wang,
Xiao \& Lei 2002, Wang et al. 2003, hereafter W02, W03; Uzdensky
2004, 2005). In the MC process energy and angular momentum are
transferred from a rotating BH to its surrounding disc. Although the
MC process cannot power jet/outflow directly, it plays an important
role in depressing disc accretion due to transfer of angular
momentum from a rotating BH to the inner disc.

In this paper we incorporate the BZ, MC and BP processes into a
model to explain the jets from AGNs and the BH binaries. Henceforth
this model is referred to as the BZ-MC-BP model. It turns out that
the three mechanisms can coexist, provided that the BH spin and the
power-law index indicating the variation of the magnetic field with
the disc radius are greater than some critical values. Since the
jets could be driven in two regimes (the BZ process in the Poynting
flux regime and the BP process in the hydromagnetic regime), this
model could be applicable to a spine/sheath jet structure observed
in BH sources of stellar and supermassive size (Meier 2003). This
paper is organized as follows. In Sect. 2 the magnetic field
configuration of the BZ-MC-BP model is outlined. The magnetic field
configuration is given based on the constraint of the screw
instability, which consists of three parts: region I, II and III
corresponding to the BZ, BP and MC processes, respectively. In Sect.
3 the condition for a centrifugally driven outflow of matter from
the disc is discussed in a parameter space. The matter outflow rate
is determined based on the radial variation of the accretion rate
due to the MC effect, and the BP power is derived based on the work
done by the magnetic torque given in BP82. It is shown that the
outer boundary radius of region II for the BP process is intimately
related to the power-law index indicating the variation of the
accretion rate in region III. In Sect. 4 we compare the relative
importance of the BZ and BP processes in jet production, and
estimate the jet powers as the sum of the BP and the BZ powers.
Finally, in Sect. 5, we discuss the potential application of the
BZ-MC-BP model to astrophysics. It turns out that the main features
of this model are consistent with the general relativistic MHD
simulations of accretion and outflow in BH systems.

 Throughout this paper the geometric units G = c = 1 are used.

\section{MAGNETIC FIELD CONFIGURATION OF BZ-MC-BP MODEL}

In order to discuss the magnetic extraction of energy and angular
momentum from BH accretion disc by virtue of the BZ, MC and BP
mechanisms, we must determine the corresponding magnetic field
configurations. The difficulty related to this issue lies in two
aspects. First, we have not enough knowledge about the origin of
magnetic field in the BH accretion disc. Some authors argued that
magnetic field could be amplified by virtue of dynamo process at
accretion disc, and the magnetic field at the BH horizon is brought
from disc accretion and kept by the magnetic pressure of the
surrounding disc (MacDonald {\&} Thorne 1982; Ghosh {\&} Abramowicz
1997; Balbus {\&} Hawley 1998). Li (2002b) and Wang et al. (2007)
discussed the origin of the magnetic field configuration
corresponding to the MC process by assuming a toroidal electric
current flowing in the equatorial plane of a Kerr BH. Unfortunately,
a reliable origin of large-scale magnetic fields in BH accretion
discs remains unclear. Second, we cannot determine the connection
between the magnetic field configurations corresponding to the BZ,
MC and BP mechanisms.

Lovelace, Romanova \& Bisnovatyi-Kogan (1995, hereafter L95)
investigated disc accretion of matter on to a rotating star with an
aligned dipole magnetic field, and they argued that when the angular
velocities of the star and disc differ substantially, the magnetic
field linking the star and disc rapidly inflates to give regions of
open field lines extending from the polar caps of the star and from
the disc. The open field line region of the disc leads to the
possibility of magnetically driven outflows.

Wang, Lei \& Ye (2006, hereafter W06) proposed a model to explain
the light curves of gamma-ray bursts by considering the effects of
screw instability of magnetic field. It turns out that the screw
instability in the BZ process (henceforth SIBZ) can coexist with the
screw instability in the MC process (henceforth SIMC), provided that
three parameters are greater than some critical values. These
parameters are (i) the BH spin defined as $a_ * \equiv J
\mathord{\left/ {\vphantom {J {M^2}}} \right.
\kern-\nulldelimiterspace} {M^2}$, (ii) the power-law index $n$
indicating the variation of the poloidal magnetic field at the disc,
$B_d^p \propto r_d ^{ - n}$ and (iii) the critical height of the
astrophysical load $H_c$  above the disc surface. The involved
quantities $M$, $J$, $r_d $ and $B_d^p $ are the BH mass, the BH
angular momentum, the disc radius and the poloidal magnetic field at
the disc, respectively.

In this paper we combine the BZ, MC and BP processes into a model by
virtue of the screw instability of the magnetic field. Based on the
constraints due to SIMC and SIBZ given in W06 we have the magnetic
field configuration related to the BZ-MC-BP model as shown in Figure
1.

\begin{figure}
\vspace{0.5cm}
\begin{center}
\includegraphics[width=7cm]{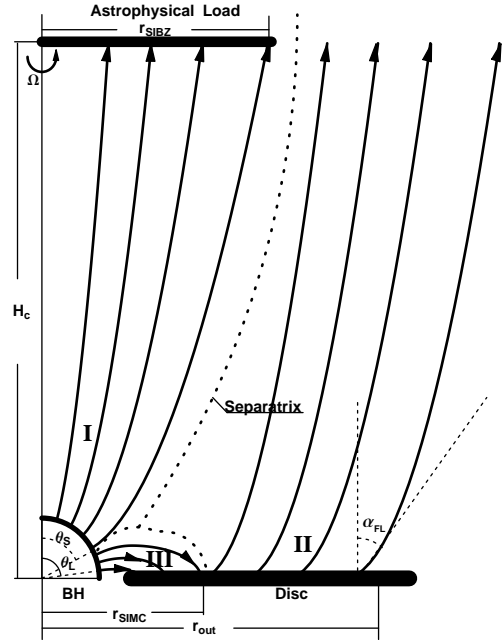}
\caption{Schematic drawing of the magnetic field configuration of
the BZ-MC-BP model} \label{fig1}
\end{center}
\end{figure}

In Figure 1  $r_{_{SIMC}} $ is the critical radius constrained by
SIMC, which is determined by

\begin{equation}
\label{eq1} {\left( {{2\pi r_{_{SIMC}} } \mathord{\left/ {\vphantom
{{2\pi r_{_{SIMC}} } {L_{MC} }}} \right. \kern-\nulldelimiterspace}
{L_{MC} }} \right)B_d^p } \mathord{\left/ {\vphantom {{\left( {{2\pi
r_{_{SIMC}} } \mathord{\left/ {\vphantom {{2\pi r_{_{SIMC}} }
{L_{MC} }}} \right. \kern-\nulldelimiterspace} {L_{MC} }}
\right)B_d^p } {B_d^T }}} \right. \kern-\nulldelimiterspace} {B_d^T
} = 1.
\end{equation}

\noindent Similarly,  $r_{_{SIBZ}} $ is the critical radius
constrained by SIBZ, which is determined by

\begin{equation}
\label{eq2} {\left( {{2\pi r_{_{SIBZ}} } \mathord{\left/ {\vphantom
{{2\pi r_{_{SIBZ}} } {L_{BZ} }}} \right. \kern-\nulldelimiterspace}
{L_{BZ} }} \right)B_L^p } \mathord{\left/ {\vphantom {{\left( {{2\pi
r_{_{SIBZ}} } \mathord{\left/ {\vphantom {{2\pi r_{_{SIBZ}} }
{L_{BZ} }}} \right. \kern-\nulldelimiterspace} {L_{BZ} }}
\right)B_L^p } {B_L^T }}} \right. \kern-\nulldelimiterspace} {B_L^T
} = 1.
\end{equation}

Equations (\ref{eq1}) and (\ref{eq2}) are derived based on the
Kruskal-Shafranov criterion: the screw instability will occur, if
the magnetic field line turns around itself about once (Kadomtsev
1966; Bateman 1978). In equation (\ref{eq1}) $L_{MC} $ is the
critical length of the poloidal field line for SIMC, and $B_d^p $
and $B_d^T $ are the poloidal and toroidal components of the
magnetic field on the disc, respectively. In equation (\ref{eq2})
$L_{BZ} $ is the critical length of the poloidal field line for
SIBZ, and $B_L^p $ and $B_L^T $ are the poloidal and toroidal
components of the magnetic field on the astrophysical load,
respectively.

As shown in Figure 1, the regions of open field lines at the horizon
and disc are referred to as regions I and II, which correspond to
the BZ and BP processes, respectively. The region of closed field
lines connecting the BH with the disc is referred to as regions III,
which corresponds to the MC process. Region I is confined to the
angular region at the horizon, $0 < \theta < \theta _S $, while
region II is confined to the radial region at the disc, $r_{_{SIMC}}
< r < r_{out} $. Regions III is confined to $\theta _S < \theta <
\theta _L $ at the horizon and to $r_{ms} < r < r_{_{SIMC}} $ at the
disc. The angle $\theta _S$  is the angular boundary between the
open and closed field lines on the horizon, and $\theta _L$  is the
lower boundary angle for the closed field lines. Throughout this
paper $\theta _L=0.45\pi$ is taken in calculations.

From Figure 1 we find that the magnetic field configuration for the
BZ-MC-BP model looks similar to that of L95 (see Figure 3 given in
L95). In both cases the open field lines extend from the central
object and from the disc, and closed field lines connect the central
object with the inner disc. However, there are several differences
between the two cases.

(1) In L95 the central object is a neutron star, and the magnetic
field lines are frozen at its surface. The open field lines are
produced by the toroidal magnetic flux which is generated out of
the closed field lines, arising from the difference of angular
velocities between the neutron star and the disc. While the
central object is a rotating BH in the BZ-MC-BP model, and the
field lines can slip on the BH horizon. The coexistence of the
open and the closed field lines at the BH horizon has been argued
in W06. In the BZ-MC-BP model we assume that the open magnetic
field at region II is amplified by the dynamo process, being
brought inwards by the disc accretion, and the critical radii
($r_{_{SIMC}} $ and $r_{_{SIBZ}} $) provide a natural constraint
to the open field lines extending from the disc to infinity as
shown in Figure 1.

(2) The directions of the open field lines across the separatrix are
opposite in L95, implying a current sheet exists along the
separatrix. A magnetic reconnection might start from this
configuration. In the BZ-MC-BP model, as shown in Figure 1, the open
field lines in region I and II are in the same direction, which can
be continuous across the separatrix, and the open magnetic field at
region II is balanced by the magnetic pressure in the regions I and
III.

(3) In L95 the magnetic field lines in region I penetrate the disc
vertically, and an outflow cannot be driven centrifugally in the BP
process. Contrary to L95, as shown in Figure 1, the poloidal
magnetic field in region II could make an angle greater than $30^0$
with the normal of the disc surface, and an outflow of matter driven
centrifugally by the open magnetic field is permitted.

Following BP82, we assume that the poloidal magnetic field in region
II varies with the disc radius $r_d $ as follows,

\begin{equation}
\label{eq3} \left( {B_d^P } \right)_{BP} \propto r_d^{{ - 5}
\mathord{\left/ {\vphantom {{ - 5} 4}} \right.
\kern-\nulldelimiterspace} 4} .
\end{equation}

\noindent Considering the balance of the magnetic pressure across
the boundary of region II and III, we have

\begin{equation}
\label{eq4} \left( {B_d^P } \right)_{BP} = B_{MC} \left( {{r_d }
\mathord{\left/ {\vphantom {{r_d } {r_{_{SIMC}} }}} \right.
\kern-\nulldelimiterspace} {r_{_{SIMC}} }} \right)^{{ - 5}
\mathord{\left/ {\vphantom {{ - 5} 4}} \right.
\kern-\nulldelimiterspace} 4} = B_{MC} \left( {\xi \mathord{\left/
{\vphantom {\xi {\xi _{SIMC} }}} \right. \kern-\nulldelimiterspace}
{\xi _{SIMC} }} \right)^{{ - 5} \mathord{\left/ {\vphantom {{ - 5}
4}} \right. \kern-\nulldelimiterspace} 4},
\end{equation}

\noindent where $B_{MC} $ is the magnetic field at $r_{_{SIMC}} $,
and $\xi \equiv {r_d } \mathord{\left/ {\vphantom {{r_d } {r_{ms}
}}} \right. \kern-\nulldelimiterspace} {r_{ms} }$ is the disc radius
in terms of $r_{ms} $.

According to W03 the magnetic field $B_{MC} $ is related to the
magnetic field at the horizon, $B_H $, by

\begin{equation}
\label{eq5} B_H 2\pi \left( {\varpi \rho } \right)_{r = r_H }
d\theta = - B_{MC} 2\pi \left( {{\varpi \rho } \mathord{\left/
{\vphantom {{\varpi \rho } {\sqrt \Delta }}} \right.
\kern-\nulldelimiterspace} {\sqrt \Delta }} \right)_{\theta = \pi
\mathord{\left/ {\vphantom {\pi 2}} \right.
\kern-\nulldelimiterspace} 2} dr_d .
\end{equation}

\noindent where $\varpi $, $\rho $ and $\Delta $ are the Kerr metric
coefficients, and they read

\begin{equation}
\label{eq61} \left\{ {\begin{array}{l}
 \varpi = \left( {\Sigma \mathord{\left/ {\vphantom {\Sigma \rho }} \right.
\kern-\nulldelimiterspace} \rho } \right)\sin \theta , \\
 \Sigma ^2 \equiv \left( {r^2 + a^2} \right)^2 - a^2\Delta \sin ^2\theta ,
\\
 \rho ^2 \equiv r^2 + a^2\cos ^2\theta , \\
 \Delta \equiv r^2 + a^2 - 2Mr. \\
 \end{array}} \right.
\end{equation}

The angular coordinate $\theta $ on the BH horizon is related to the
dimensionless disc radius $\xi $ by the following mapping relation:

\begin{equation}
\label{eq71} \cos \theta - \cos \theta _L = \int_1^\xi
{\mbox{G}\left( {a_ * ;\xi ,n} \right)d\xi } ,
\end{equation}

\noindent where the function $\mbox{G}\left( {a_ * ;\xi ,n} \right)$
has been given in W03 as follows,

\begin{equation}
\label{eq81}\begin{array}{l}
 \mbox{G}\left( {a_ * ;\xi ,n} \right)=\\
\frac{\xi ^{1 - n}\chi _{ms}^2 \sqrt {1 + a_ * ^2 \chi _{ms}^{ -
4} \xi ^{ - 2} + 2a_ * ^2 \chi _{ms}^{ - 6} \xi ^{ - 3}} }{2\sqrt
{\left( {1 + a_
* ^2 \chi _{ms}^{ - 4} + 2a_ * ^2 \chi _{ms}^{ - 6} }
\right)\left( {1 - 2\chi _{ms}^{ - 2} \xi ^{ - 1} + a_ * ^2 \chi
_{ms}^{ - 4} \xi ^{ - 2}} \right)} }.
\end{array}
\end{equation}

In equation (\ref{eq81}) $\chi _{ms} $ is defined in terms of the
radius of the innermost stable circular orbit, $r_{ms} = M\chi
_{ms}^2 $, which is regarded as the inner edge of the disc.
Incorporating equations (\ref{eq5}) and (\ref{eq71}), we have


\begin{equation}
\label{eq7}\begin{array}{l}B_{MC} =\\ B_H \left. {\frac{2\left( {1
+ q} \right)\sqrt {1 + a_ * ^2 \xi ^{ - 2}\chi _{ms}^{ - 4} - 2\xi
^{ - 1}\chi _{ms}^{ - 2} } }{\xi \chi _{ms}^4 \sqrt {1 + a_ * ^2
\xi ^{ - 2}\chi _{ms}^{ - 4} + 2a_ * ^2 \xi ^{ - 3}\chi _{ms}^{ -
6} } }\mbox{G}\left( {a_ * ;\xi ,n} \right)} \right|_{\xi = \xi
_{SIMC}}.
\end{array}
\end{equation}

\noindent where $q \equiv \sqrt {1 - a_ * ^2 } $ is a function of
the BH spin $a_ *$. Thus we can determine the poloidal magnetic
field in region II by combining equation (\ref{eq4}) with equation
(\ref{eq7}).

\section{OUTFLOW RATE, POWER AND TORQUE IN BP PROCESS}

\subsection{Condition for outflow driven in hydromagnetic regime}

As argued in BP82, an outflow of matter can be launched
centrifugally, only if the angle of the poloidal component of the
magnetic field with the normal of the disc is greater than a
critical value, i.e., $\alpha _{FL} > 30^0$. We can discuss the
condition for driving the outflow based on the magnetic
configurations given in Figure 1, and the angle $\alpha _{FL} $ can
be estimated by

\begin{equation}
\label{eq101} \alpha _{FL} = \tan ^{ - 1}\left[ {{\left( {\xi
_{SIBZ} - \xi _{SIMC} } \right)\chi _{ms}^2 } \mathord{\left/
{\vphantom {{\left( {\xi _{SIBZ} - \xi _{SIMC} } \right)\chi _{ms}^2
} {h_c }}} \right. \kern-\nulldelimiterspace} {h_c }} \right],
\end{equation}

\noindent where $h_c \equiv {H_c } \mathord{\left/ {\vphantom {{H_c
} M}} \right. \kern-\nulldelimiterspace} M$ is the dimensionless
height of the load above the disc surface.

Incorporating equation (\ref{eq101}) with the criterions (\ref{eq1})
and (\ref{eq2}), we obtain the values of $\alpha _{FL} $
corresponding to the concerned parameters of SIBZ and SIMC as shown
in Table 1, in which we have $\alpha _{FL} > 30^0$ for driving the
outflow in the hydromagnetic regime.


\begin{table*}
\caption{The angle $\alpha _{FL} $ corresponding to the parameters
of SIBZ and SIMC }
\begin{tabular}
{|p{50pt}|p{53pt}|p{64pt}|p{64pt}|p{64pt}|p{64pt}|} \hline $a_ *
$& $n$ & $h_c $&  $\xi _{SIBZ} $& $\xi _{SIMC} $&
$\alpha_{_{FL}}$ \\
\hline 0.40& 7.6& 52.41&  12.67& 2.20&
42.7$^{0}$ \\
\hline 0.50& 7.1& 41.55&  10.64& 2.21&
40.7$^{0}$ \\
\hline 0.60& 6.6& 33.73&  9.31& 2.22&
38.8$^{0}$ \\
\hline 0.70& 6.1& 27.40&  8.39& 2.23&
37.3$^{0}$ \\
\hline 0.80& 5.6& 21.55&  7.74& 2.22&
36.7$^{0}$ \\
\hline
\end{tabular}
\label{tab1}
\end{table*}

It is shown in Table 1 that the angle $\alpha _{FL} $ can be
determined by three parameters, $a_ * $, $n$ and $h_c $, and the
correlation of $\alpha _{FL} $ with the three parameters is depicted
in the $a_ * - n$ parameter space for the given $h_c $ in Figure 2.
The coexistence of SIBZ and SIMC is indicated by the shaded regions,
which are bounded by two contours, i.e., $\theta _S = 0$ and $h_c =
$52.41, 41.55 and 33.73 in Figure 2a, 2b and 2c, respectively. Each
shaded region is further divided by the contour of $\alpha _{FL} =
30^0$, below and above which we have $\alpha _{FL} < 30^0$ and
$\alpha _{FL} > 30^0$ in regions A and B, respectively.

Thus the two regimes of outflows driven can be determined by regions
A and B, which correspond to the Poynting flux regime and
hydromagnetic regime, respectively. In this paper we confine the
discussion to the BP process by taking the values of $a_ * $ and $n$
in region B.

\begin{figure*}
\begin{center}
{\includegraphics[width=5.0cm]{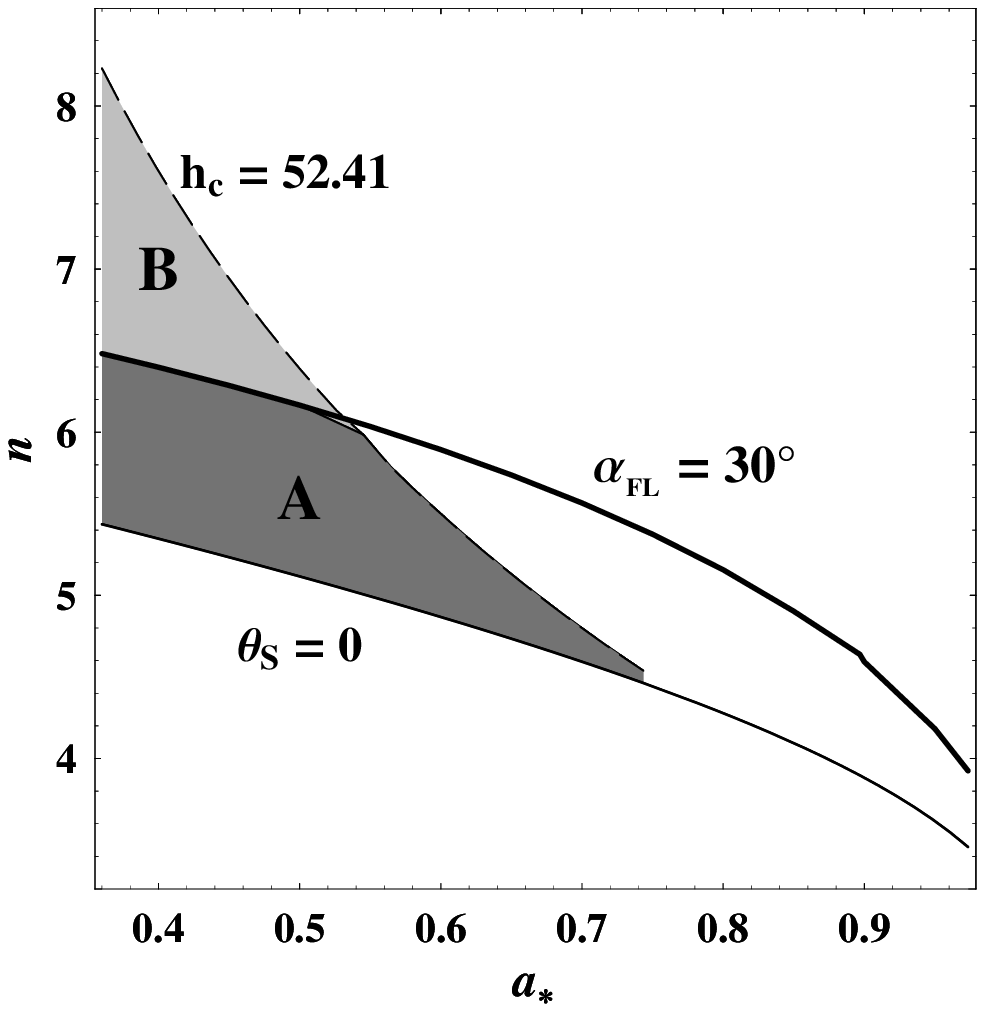} \hfill
\includegraphics[width=5.0cm]{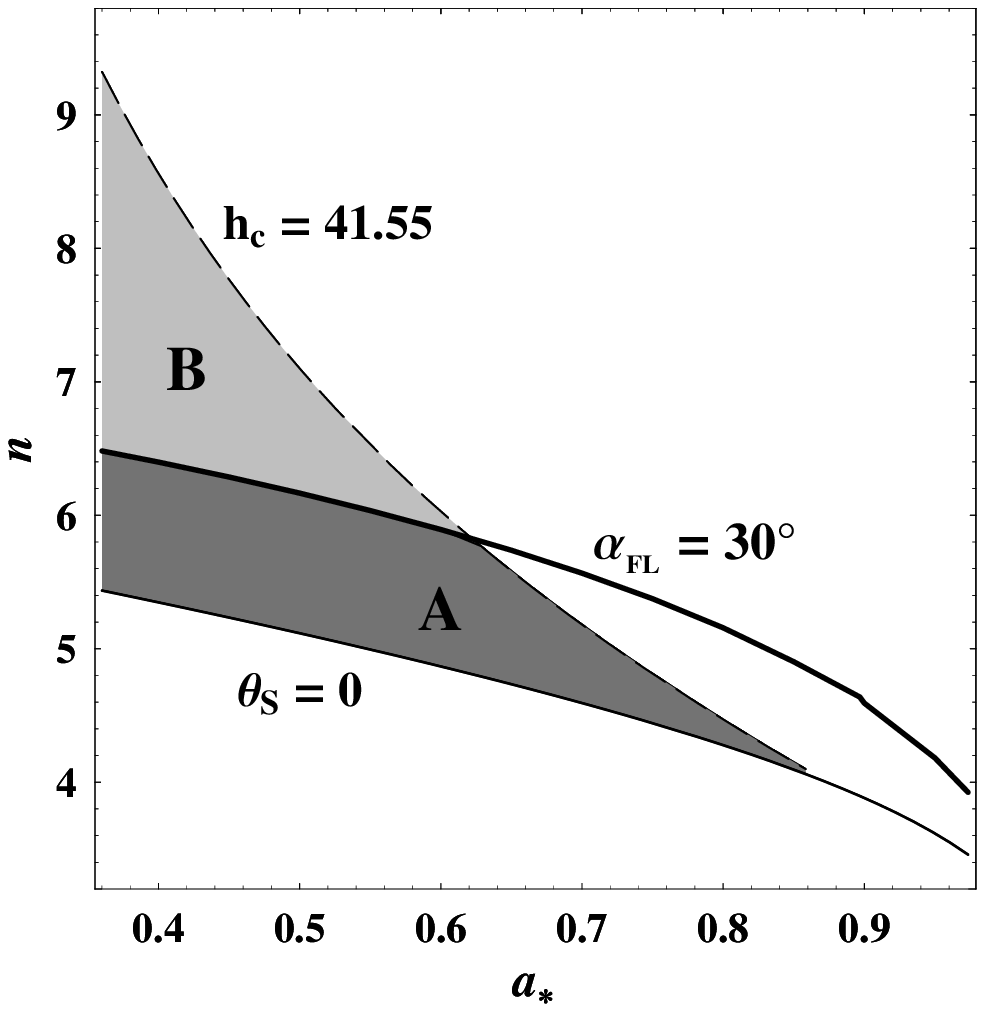} \hfill
\includegraphics[width=5.0cm]{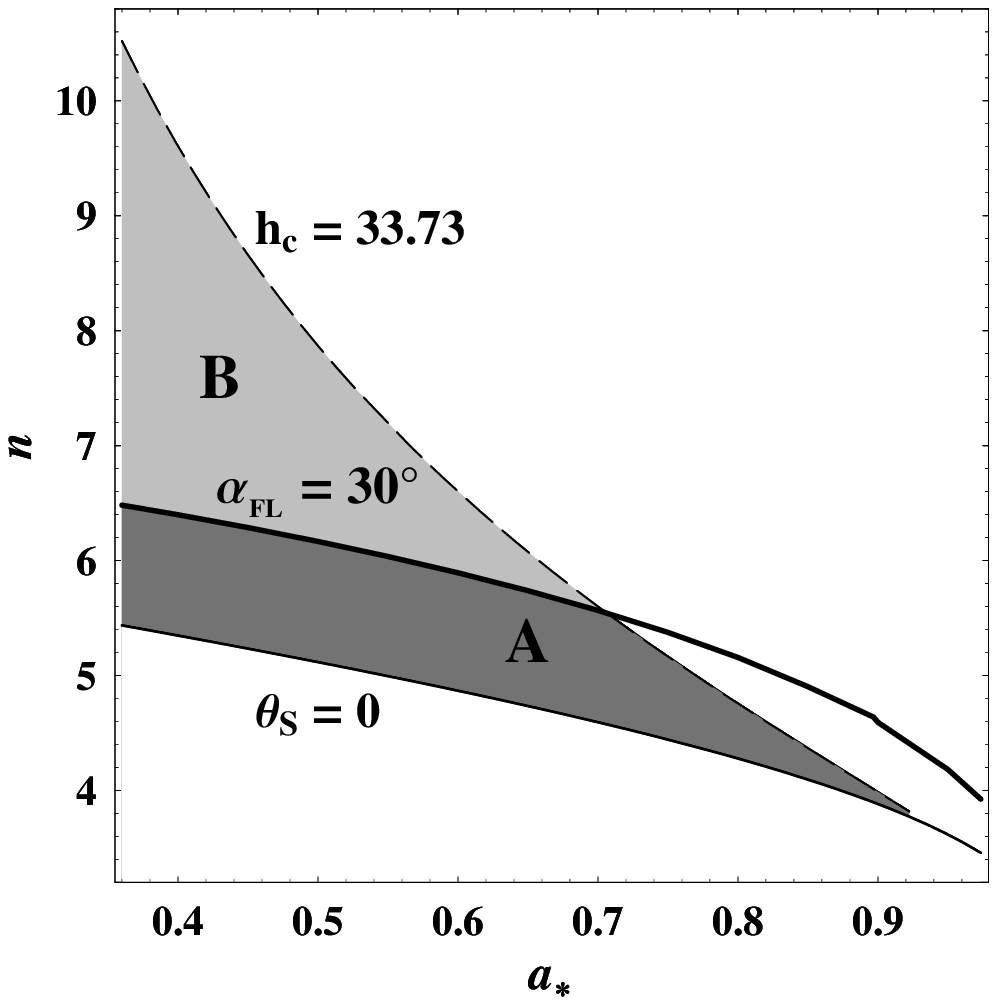}
\centerline{\hspace{0.6cm}(a)\hspace{6cm}(b)\hspace{6cm}(c)} }
\caption{The shaded regions between the contour of $\theta_{S}$=0
for (solid line) and the contours of $h_{c} $(dashed lines) are
divided by the contour of $\alpha _{FL}$=30$^{0}$ (thick solid line)
for $h_{c} $ = 52.41, 41.55 and 33.73 in Figures 2a, 2b and 2c,
respectively. Regions A and B indicate the parameters for $\alpha
_{FL}$less and greater than 30$^{0}$, respectively.} \label{fig2}
\end{center}
\end{figure*}

\subsection{Rate of matter outflow and BP power and torque}

Since angular momentum is transferred magnetically from a rotating
BH to the inner disc in the MC process, the accretion rate is
probably depressed. Thus we assume that the accretion rate in region
II obeys the following relation,

\begin{equation}
\label{eq8} {\dot {M}_d } \mathord{\left/ {\vphantom {{\dot {M}_d }
{dr_d }}} \right. \kern-\nulldelimiterspace} {dr_d } > 0,  \quad for
 \quad\quad r_{_{SIMC}} < r_d < r_{out} . \end{equation}

\noindent Based on mass conservation the outflow rate is given by

\begin{equation}
\label{eq9} \dot {M}_{outflow} = \left( {\dot {M}_d } \right)_{out}
- \left( {\dot {M}_d } \right)_{MC} ,
\end{equation}

\noindent where $\left( {\dot {M}_d } \right)_{MC} $ and $\left(
{\dot {M}_d } \right)_{out} $ are the accretion rates at
$r_{_{SIMC}} $ and $r_{out} $, respectively.

As the magnetic field on the BH horizon is supported by the
surrounding disc, there are some relations between $B_H $ and $\dot
{M}_d $. One of them is given by considering the balance between the
magnetic pressure on the horizon and the ram pressure of the
innermost parts of an accretion flow (Moderski, Sikora {\&} Lasota
1997), i.e.,

\begin{equation}
\label{eq10} {B_H^2 } \mathord{\left/ {\vphantom {{B_H^2 } {\left(
{8\pi } \right)}}} \right. \kern-\nulldelimiterspace} {\left( {8\pi
} \right)} = P_{ram} \sim \rho c^2\sim {\dot {M}_d } \mathord{\left/
{\vphantom {{\dot {M}_d } {\left( {4\pi r_H^2 } \right)}}} \right.
\kern-\nulldelimiterspace} {\left( {4\pi r_H^2 } \right)},
\end{equation}

As a simple analysis, we take the accretion rate in region III as a
constant, which is related to the magnetic field at BH horizon by
the following relation,

\begin{equation}
\label{eq11} \left( {\dot {M}_d } \right)_{MC} = \alpha _m B_H^2
r_H^2
\end{equation}

\noindent In equation (\ref{eq11}) the parameter $\alpha _m $ is a
adjustable parameter due to the uncertainty of equation
(\ref{eq10}), and we take $\alpha _m = 0.1$ in calculations.

Assuming that the accretion rate in region II varies with disc
radius in a power-law, $\dot {M}_d \propto r_d^S $, we obtain its
expression by combining equation (\ref{eq11}) as follows,


\begin{equation}
\label{eq12} \dot {M}_d = \alpha _m B_H^2 r_H^2 \left( {{r_d }
\mathord{\left/ {\vphantom {{r_d } {r_{_{SIMC}} }}} \right.
\kern-\nulldelimiterspace} {r_{_{SIMC}} }} \right)^S \quad for
 \quad r_{_{SIMC}} < r_d < r_{out}  \end{equation}

\noindent or


\begin{equation}
\label{eq13}\begin{array}{l}\dot {m}_d \equiv {\dot {M}_d }
\mathord{\left/ {\vphantom {{\dot {M}_d } {B_H^2 M^2}}} \right.
\kern-\nulldelimiterspace} {B_H^2 M^2} = \alpha _m \left( {1 + q}
\right)^2\left( {\xi \mathord{\left/ {\vphantom {\xi {\xi _{SIMC}
}}} \right. \kern-\nulldelimiterspace} {\xi _{SIMC} }} \right)^S
\\ \quad\quad\quad\quad\quad\quad\quad\quad\quad\quad\quad\quad\quad for \quad \xi _{SIMC} < \xi < \xi _{out}  \end{array}
\end{equation}

\noindent where $B_H^2 M^2 \equiv \left( {B_4^2 m_H^2 \times
7.32\times 10^7} \right)g \cdot s^{ - 1}$.

Thus the outflow rate from a ring with width $r_d - r_d + dr_d $ can
be expressed as

\begin{equation}
\label{eq14} d\dot {m}_{outflow} = \alpha _m \left( {1 + q}
\right)^2S\left( {\xi \mathord{\left/ {\vphantom {\xi {\xi _{SIMC}
}}} \right. \kern-\nulldelimiterspace} {\xi _{SIMC} }} \right)^{S -
1}d\left( {\xi \mathord{\left/ {\vphantom {\xi {\xi _{SIMC} }}}
\right. \kern-\nulldelimiterspace} {\xi _{SIMC} }} \right),
\end{equation}

\noindent and the total outflow rate driven from region II is

\begin{equation}
\label{eq15} \dot {m}_{outflow} = \alpha _m \left( {1 + q}
\right)^2\left( {\varsigma _{out}^S - 1} \right),
\end{equation}

\noindent where $\dot {m}_{outflow} \equiv {\dot {M}_{outflow} }
\mathord{\left/ {\vphantom {{\dot {M}_{outflow} } {B_H^2 M^2}}}
\right. \kern-\nulldelimiterspace} {B_H^2 M^2}$ and $\varsigma
_{out} \equiv {\xi _{out} } \mathord{\left/ {\vphantom {{\xi _{out}
} {\xi _{SIMC} }}} \right. \kern-\nulldelimiterspace} {\xi _{SIMC}
}$.

According to BP82 the specific energy $e$ and angular momentum $l$
are constants along each open field line, and they read

\begin{equation}
\label{eq151} e = {\upsilon ^2} \mathord{\left/ {\vphantom
{{\upsilon ^2} 2}} \right. \kern-\nulldelimiterspace} 2 + h + \Phi -
{\omega rB_\varphi } \mathord{\left/ {\vphantom {{\omega rB_\varphi
} k}} \right. \kern-\nulldelimiterspace} k, \quad l = r\upsilon
_\varphi - {rB_\varphi } \mathord{\left/ {\vphantom {{rB_\varphi }
k}} \right. \kern-\nulldelimiterspace} k.
\end{equation}

The quantities involved in equation (\ref{eq151}) are interpreted as
follows. The quantity $r$ is the cylindrical radius, and $\omega $
is the angular velocity of the foot point of the field line, being
taken as Keplerian angular velocity of a thin disc and remaining
constant along the field line. The quantities $h$ and $\Phi $ are
respectively the enthalpy per unit mass and gravitational potential,
and the quantities $\upsilon $ and $\upsilon _\varphi $ are the
velocity of the streaming gas and its toroidal component,
respectively. The parameter $k$ is defined as

\begin{equation}
\label{eq161} k \mathord{\left/ {\vphantom {k {4\pi }}} \right.
\kern-\nulldelimiterspace} {4\pi } = {\rho _m \upsilon _P }
\mathord{\left/ {\vphantom {{\rho _m \upsilon _P } {B_P }}} \right.
\kern-\nulldelimiterspace} {B_P },
\end{equation}

\noindent which is interpreted as the ratio of the mass flux to the
magnetic flux, and it also remains constant along each field line.
The quantities $\rho _m $ and $\upsilon _P $ are the mass density
and the poloidal velocity of the streaming gas, respectively. The
toroidal velocity $\upsilon _\varphi $ is related to the toroidal
magnetic field $B_\varphi $ by

\begin{equation}
\label{eq171} \upsilon _\varphi = {\upsilon _P B_\varphi }
\mathord{\left/ {\vphantom {{\upsilon _P B_\varphi } {B_P }}}
\right. \kern-\nulldelimiterspace} {B_P } + r\omega ,
\end{equation}

\noindent where ${\upsilon _P B_\varphi } \mathord{\left/ {\vphantom
{{\upsilon _P B_\varphi } {B_P }}} \right.
\kern-\nulldelimiterspace} {B_P }$ is the toroidal velocity of the
streaming gas with respect to the rotating field line, and $r\omega
$ is the toroidal velocity of the field line itself. Incorporating
equations (\ref{eq151})---(\ref{eq171}), we have

\begin{equation}
\label{eq181} B_\varphi = \frac{k\left( {r\omega - l \mathord{\left/
{\vphantom {l r}} \right. \kern-\nulldelimiterspace} r} \right)}{1 -
\left( {{\upsilon _P } \mathord{\left/ {\vphantom {{\upsilon _P }
{\upsilon _{AP} }}} \right. \kern-\nulldelimiterspace} {\upsilon
_{AP} }} \right)^2} = \frac{kr\omega \left( {1 - {\lambda r_d^2 }
\mathord{\left/ {\vphantom {{\lambda r_d^2 } {r^2}}} \right.
\kern-\nulldelimiterspace} {r^2}} \right)}{1 - \left( {{\upsilon _P
} \mathord{\left/ {\vphantom {{\upsilon _P } {\upsilon _{AP} }}}
\right. \kern-\nulldelimiterspace} {\upsilon _{AP} }} \right)^2},
\end{equation}

\noindent where $\upsilon _{AP} \equiv {B_P } \mathord{\left/
{\vphantom {{B_P } {\sqrt {4\pi \rho } }}} \right.
\kern-\nulldelimiterspace} {\sqrt {4\pi \rho } }$ is the poloidal
component of the Alfven velocity, and $\lambda \equiv l
\mathord{\left/ {\vphantom {l {r_d^2 \omega }}} \right.
\kern-\nulldelimiterspace} {r_d^2 \omega }$ is defined as the ratio
of the specific angular momentum to that at the midplane of the disc
in BP82.

Incorporating equations (\ref{eq151})---(\ref{eq181}), we have

\begin{equation}
\label{eq19} \upsilon _\varphi = r\omega \left[ {1 + \frac{\left(
{{\upsilon _P } \mathord{\left/ {\vphantom {{\upsilon _P } {\upsilon
_{AP} }}} \right. \kern-\nulldelimiterspace} {\upsilon _{AP} }}
\right)^2\left( {1 - {\lambda r_d^2 } \mathord{\left/ {\vphantom
{{\lambda r_d^2 } {r^2}}} \right. \kern-\nulldelimiterspace} {r^2}}
\right)}{1 - \left( {{\upsilon _P } \mathord{\left/ {\vphantom
{{\upsilon _P } {\upsilon _{AP} }}} \right.
\kern-\nulldelimiterspace} {\upsilon _{AP} }} \right)^2}} \right].
\end{equation}

According to BP82 the term $ - {\omega rB_\varphi } \mathord{\left/
{\vphantom {{\omega rB_\varphi } k}} \right.
\kern-\nulldelimiterspace} k$ in equation (\ref{eq151}) represents
the work done on the streaming gas by the magnetic torque. Combining
equation (\ref{eq181}) with the work of the magnetic torque and the
outflow rate, we derive the BP power and torque as follows,

\begin{equation}
\label{eq201} \begin{array}{l} dP_{BP} = \left( { - {\omega
rB_\varphi } \mathord{\left/ {\vphantom {{\omega rB_\varphi } k}}
\right. \kern-\nulldelimiterspace} k} \right)d\dot {M}_{outflow}
\\
\ \ \ \ \ \ \ \ = \frac{\omega ^2\left( {\lambda r_d^2 - r^2}
\right)}{1 - \left( {{\upsilon _P } \mathord{\left/ {\vphantom
{{\upsilon _P } {\upsilon _{AP} }}} \right.
\kern-\nulldelimiterspace} {\upsilon _{AP} }} \right)^2}d\dot
{M}_{outflow} ,
\end{array}
\end{equation}

\begin{equation}
\label{eq211} dT_{BP} = {dP_{BP} } \mathord{\left/ {\vphantom
{{dP_{BP} } \omega }} \right. \kern-\nulldelimiterspace} \omega =
\frac{\omega \left( {\lambda r_d^2 - r^2} \right)}{1 - \left(
{{\upsilon _P } \mathord{\left/ {\vphantom {{\upsilon _P } {\upsilon
_{AP} }}} \right. \kern-\nulldelimiterspace} {\upsilon _{AP} }}
\right)^2}d\dot {M}_{outflow} .
\end{equation}

In order to avoid the infinite BP power and torque at the Alfven
surface as $\upsilon _P \to \upsilon _{AP} $, we have

\begin{equation}
\label{eq221} \lambda = \left( {{r_A } \mathord{\left/ {\vphantom
{{r_A } {r_d }}} \right. \kern-\nulldelimiterspace} {r_d }}
\right)^2.
\end{equation}

And equations (\ref{eq201}) and (\ref{eq211}) can be rewritten as

\begin{equation}
\label{eq23} dP_{BP} = \frac{\omega ^2\left( {r_A^2 - r^2}
\right)}{1 - \left( {{\upsilon _P } \mathord{\left/ {\vphantom
{{\upsilon _P } {\upsilon _{AP} }}} \right.
\kern-\nulldelimiterspace} {\upsilon _{AP} }} \right)^2}d\dot
{M}_{outflow} ,
\end{equation}

\begin{equation}
\label{eq24} dT_{BP} = {dP_{BP} } \mathord{\left/ {\vphantom
{{dP_{BP} } \omega }} \right. \kern-\nulldelimiterspace} \omega =
\frac{\omega \left( {r_A^2 - r^2} \right)}{1 - \left( {{\upsilon
_P } \mathord{\left/ {\vphantom {{\upsilon _P } {\upsilon _{AP}
}}} \right. \kern-\nulldelimiterspace} {\upsilon _{AP} }}
\right)^2}d\dot {M}_{outflow} .
\end{equation}

Since the velocity of the gas increases with the cylinder radius in
approaching the Alfven velocity, we assume that $\upsilon _P $
varies with $r$ as follows,

\begin{equation}
\label{eq25} {\upsilon _P } \mathord{\left/ {\vphantom {{\upsilon _P
} {\upsilon _{AP} }}} \right. \kern-\nulldelimiterspace} {\upsilon
_{AP} } = \left( {r \mathord{\left/ {\vphantom {r {r_A }}} \right.
\kern-\nulldelimiterspace} {r_A }} \right)^\alpha ,
\end{equation}

\noindent where $\alpha $ is a parameter to be determined.
Substituting equation (29) into equation (27) and taking limit for
$r \to r_A $ by using L'Hospital law, we have

\begin{equation}
\label{eq26} dP_{BP} = \mathop {\lim }\limits_{r \to r_A }
\frac{\omega ^2\left( {r_A^2 - r^2} \right)}{1 - \left( {r
\mathord{\left/ {\vphantom {r {r_A }}} \right.
\kern-\nulldelimiterspace} {r_A }} \right)^{2\alpha }}d\dot
{M}_{outflow} = \frac{\omega ^2r_A^2 }{\alpha }d\dot {M}_{outflow} .
\end{equation}

Considering the fact that $B_\varphi $ is dominative over $B_P $
near the Alfven surface, we infer that $\upsilon _P < < \upsilon
_\varphi $ and take the specific kinetic energy of the streaming gas
as ${\omega ^2r_A^2 } \mathord{\left/ {\vphantom {{\omega ^2r_A^2 }
\alpha }} \right. \kern-\nulldelimiterspace} \alpha $ with $\alpha =
2$ in equation (30). Substituting equation (26) into equation (30),
we express the BP power and torque at the Alfven surface as follows,

\begin{equation}
\label{eq27} dP_{BP} = \left( {\lambda \mathord{\left/ {\vphantom
{\lambda 2}} \right. \kern-\nulldelimiterspace} 2} \right)\omega
^2r_d^2 d\dot {M}_{outflow} ,
\end{equation}

\begin{equation}
\label{eq28} dT_{BP} = \left( {\lambda \mathord{\left/ {\vphantom
{\lambda 2}} \right. \kern-\nulldelimiterspace} 2} \right)\omega
r_d^2 d\dot {M}_{outflow} .
\end{equation}

Incorporating equation (17) and integrating equations (31) and (32)
from $\xi _{SIMC} $ to $\xi _{out} $, we have

\begin{equation}
\label{eq29} \tilde {P}_{BP} = \left( {{\lambda \alpha _m }
\mathord{\left/ {\vphantom {{\lambda \alpha _m } 2}} \right.
\kern-\nulldelimiterspace} 2} \right)\frac{\left( {1 + q}
\right)^2}{\xi _{SIMC} \chi _{ms}^2 }\frac{S\left( {\varsigma
_{out}^{S - 1} - 1} \right)}{S - 1},
\end{equation}

\begin{equation}
\label{eq30} \tilde {T}_{BP} = \left( {{\lambda \alpha _m }
\mathord{\left/ {\vphantom {{\lambda \alpha _m } 2}} \right.
\kern-\nulldelimiterspace} 2} \right)\left( {1 + q} \right)^2\chi
_{ms} \xi _{SIMS}^{1 \mathord{\left/ {\vphantom {1 2}} \right.
\kern-\nulldelimiterspace} 2} \frac{S\left( {\varsigma _{out}^{S + 1
\mathord{\left/ {\vphantom {1 2}} \right. \kern-\nulldelimiterspace}
2} - 1} \right)}{S + 1 \mathord{\left/ {\vphantom {1 2}} \right.
\kern-\nulldelimiterspace} 2},
\end{equation}

\noindent where $\tilde {P}_{BP} \equiv {P_{BP} } \mathord{\left/
{\vphantom {{P_{BP} } {P_0 }}} \right. \kern-\nulldelimiterspace}
{P_0 }$, $\tilde {T}_{BP} \equiv {T_{BP} } \mathord{\left/
{\vphantom {{T_{BP} } {T_0 }}} \right. \kern-\nulldelimiterspace}
{T_0 }$ and

\begin{equation}
\label{eq31} \left\{ {\begin{array}{l}
 P_0 \equiv \left( {B_H^p } \right)^2M^2 \approx B_4^2 m_H^2 \times
6.59\times 10^{28}erg \cdot s^{ - 1} \\
 T_0 = \left\langle {B_H^2 } \right\rangle M^3 \approx B_4^2 m_H^3 \times
3.26\times 10^{23}g \cdot cm^2 \cdot s^{ - 2} \\
 \end{array}} \right.
\end{equation}

The powers and torques in the BZ and MC processes have been derived
in W02 as follows,

\begin{equation}
\label{eq32} \tilde {P}_{BZ} \equiv {P_{BZ} } \mathord{\left/
{\vphantom {{P_{BZ} } {P_0 }}} \right. \kern-\nulldelimiterspace}
{P_0 } = 2a_ * ^2 \int_0^{\theta _S } {\frac{k\left( {1 - k}
\right)\sin ^3\theta d\theta }{2 - \left( {1 - q} \right)\sin
^2\theta }} ,
\end{equation}

\begin{equation}
\label{eq33} \tilde {T}_{BZ} \equiv {T_{BZ} } \mathord{\left/
{\vphantom {{T_{BZ} } {T_0 }}} \right. \kern-\nulldelimiterspace}
{T_0 } = 4a_ * \left( {1 + q} \right)\int_0^{\theta _S }
{\frac{\left( {1 - k} \right)\sin ^3\theta d\theta }{2 - \left( {1 -
q} \right)\sin ^2\theta }} ,
\end{equation}

\begin{equation}
\label{eq34} \tilde {P}_{MC} \equiv {P_{MC} } \mathord{\left/
{\vphantom {{P_{MC} } {P_0 }}} \right. \kern-\nulldelimiterspace}
{P_0 } = 2a_ * ^2 \int_{\theta _S }^{\theta _L } {\frac{\beta \left(
{1 - \beta } \right)\sin ^3\theta d\theta }{2 - \left( {1 - q}
\right)\sin ^2\theta }} ,
\end{equation}

\begin{equation}
\label{eq35} \tilde {T}_{MC} \equiv {T_{MC} } \mathord{\left/
{\vphantom {{T_{MC} } {T_0 }}} \right. \kern-\nulldelimiterspace}
{T_0 } = 4a_ * \left( {1 + q} \right)\int_{\theta _S }^{\theta _L }
{\frac{\left( {1 - \beta } \right)\sin ^3\theta d\theta }{2 - \left(
{1 - q} \right)\sin ^2\theta }} ,
\end{equation}

\noindent where $k$ and $\beta $ are the ratios of the angular
velocity of the field line to that of the BH horizon in the BZ and
MC processes, respectively.

Now we are going to discuss the relation between the power-law index
S and the ratio $\varsigma _{out} $ based on the transfer of angular
momentum from region III to region II.

\begin{table*}
\caption{The values of the derived parameters with $\xi _{out} / \xi
_{SIMC}  = 10^2$}
\begin{tabular}
{|p{50pt}|p{53pt}|p{64pt}|p{64pt}|p{64pt}|p{64pt}|p{64pt}|} \hline
$a_ * $& $n$ & $h_c $& $\xi _{SIBZ} $& $\xi _{SIMC} $& $S$ \\
\hline 0.40& 7.6& 52.41&  12.67& 2.20&
0.0353 \\
\hline 0.50& 7.1& 41.55&  10.64& 2.21&
0.0355 \\
\hline 0.60& 6.6& 33.73&  9.31& 2.22&
0.0358 \\
\hline 0.70& 6.1& 27.40&  8.39& 2.23&
0.0362 \\
\hline 0.80& 5.6& 21.55&  7.74& 2.22&
0.0366 \\
\hline
\end{tabular}
\label{tab2}
\end{table*}

\begin{figure*}
\begin{center}
{\includegraphics[width=5.0cm]{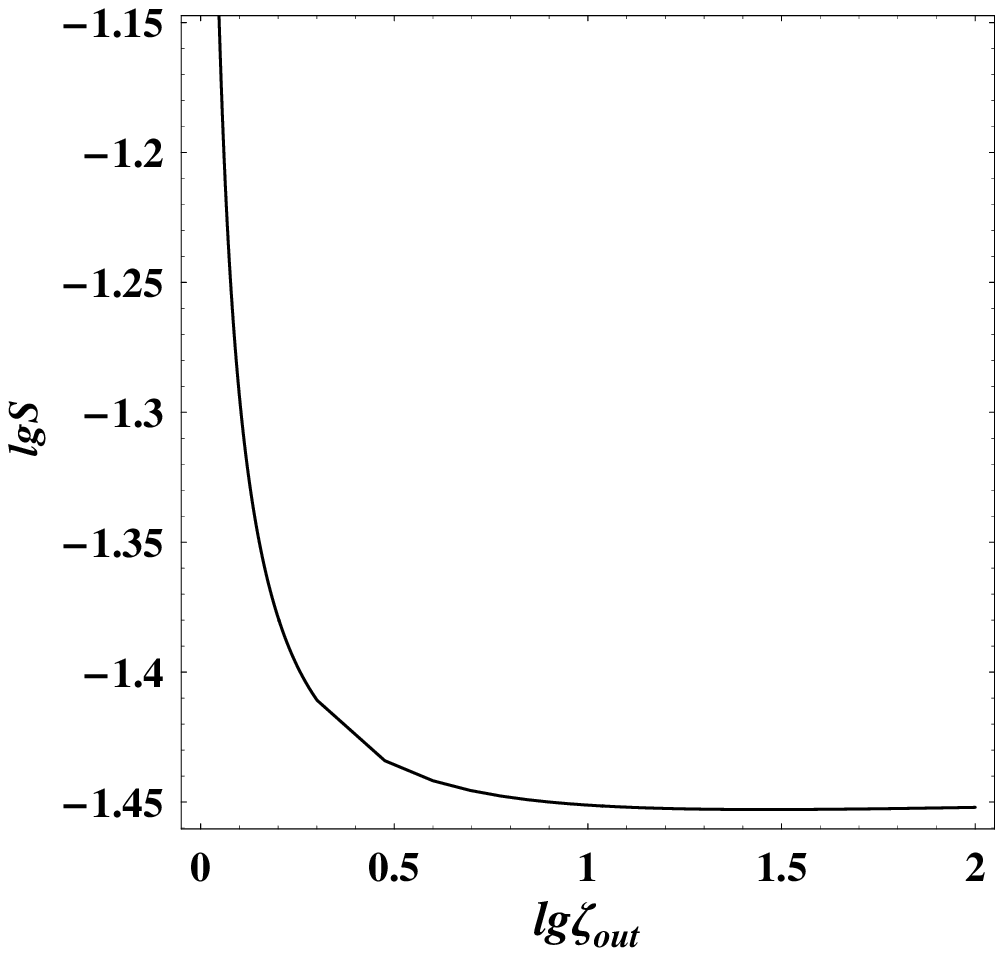} \hfill
\includegraphics[width=5.0cm]{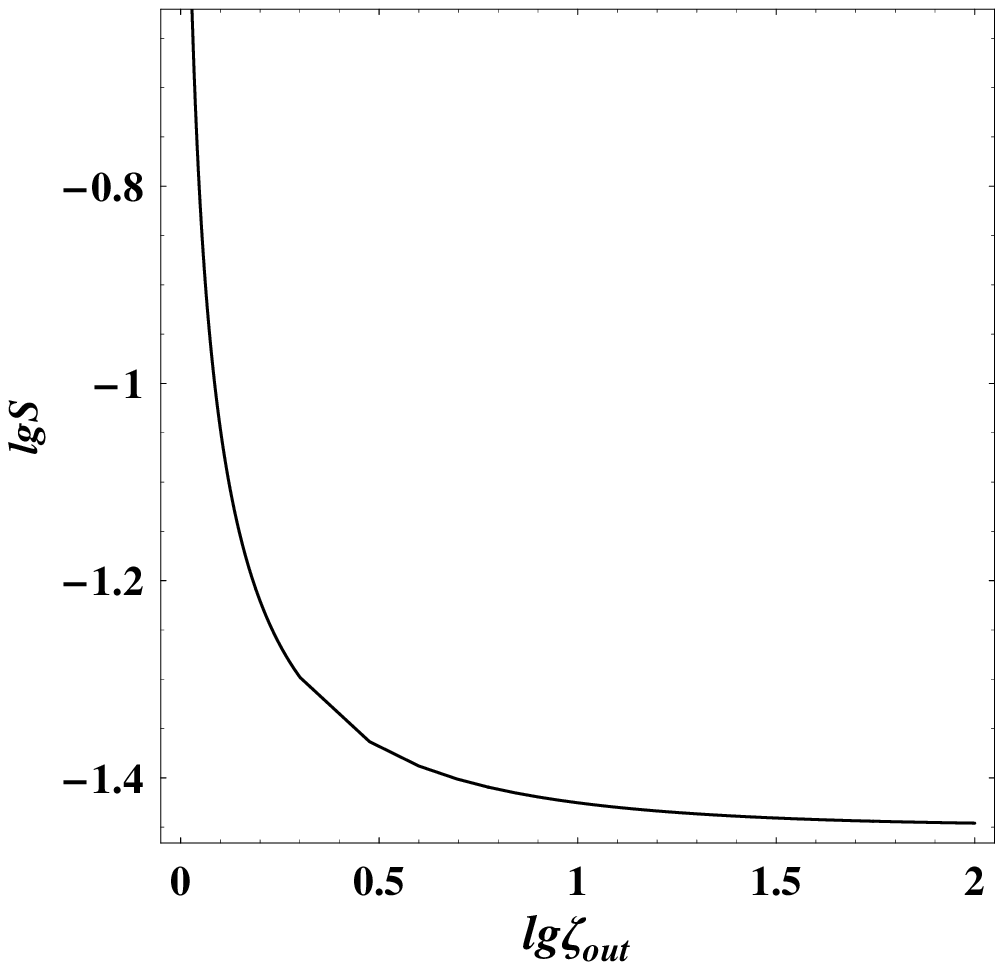} \hfill
\includegraphics[width=5.0cm]{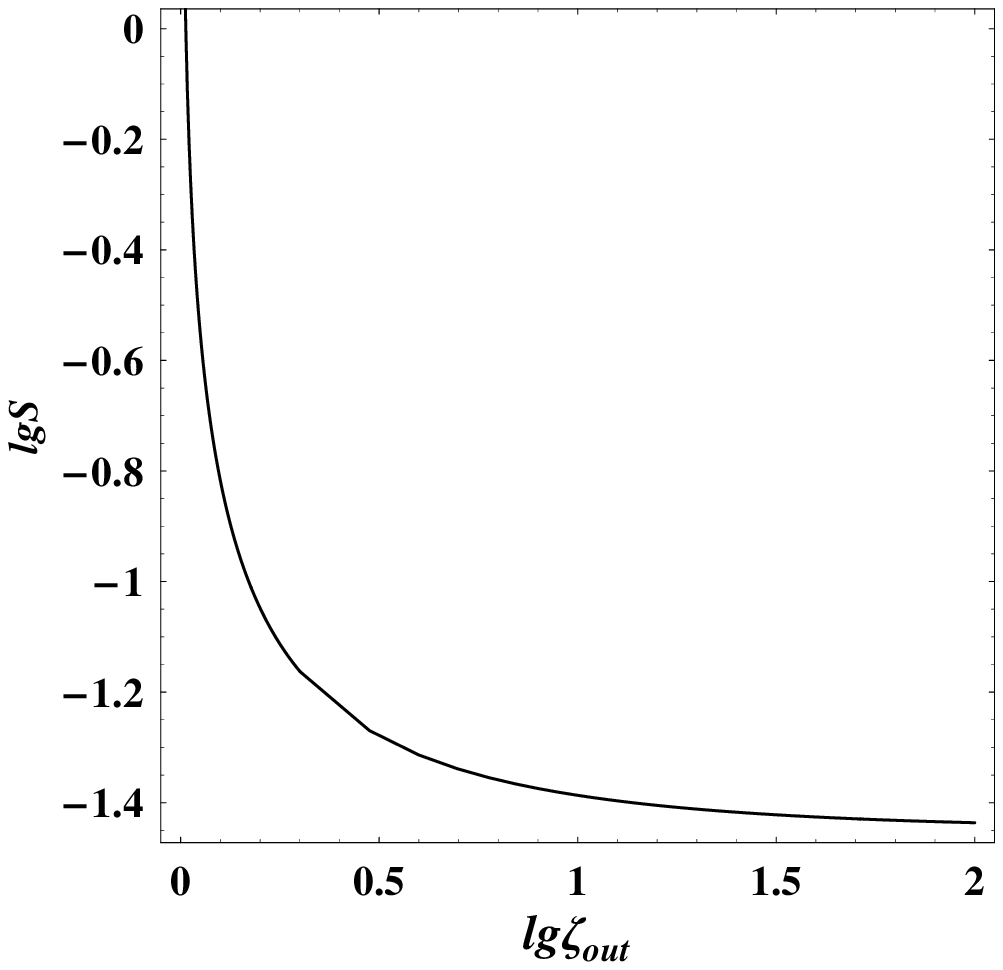}
\centerline{\hspace{0.6cm}(a)\hspace{6cm}(b)\hspace{6cm}(c)} }
\caption{The curves of $lg S $ versus $lg \varsigma _{out} $ for
different values of $a_{*}$ and $n$, (a) $a_ * = 0.40$, $n = 7.6$
(b) $a_ * = 0.60$, $n = 6.6$ and (c) $a_* = 0.80$, $n = 5.6$.}
\label{fig3}
\end{center}
\end{figure*}

\begin{figure*}
\begin{center}
{\includegraphics[width=5.0cm]{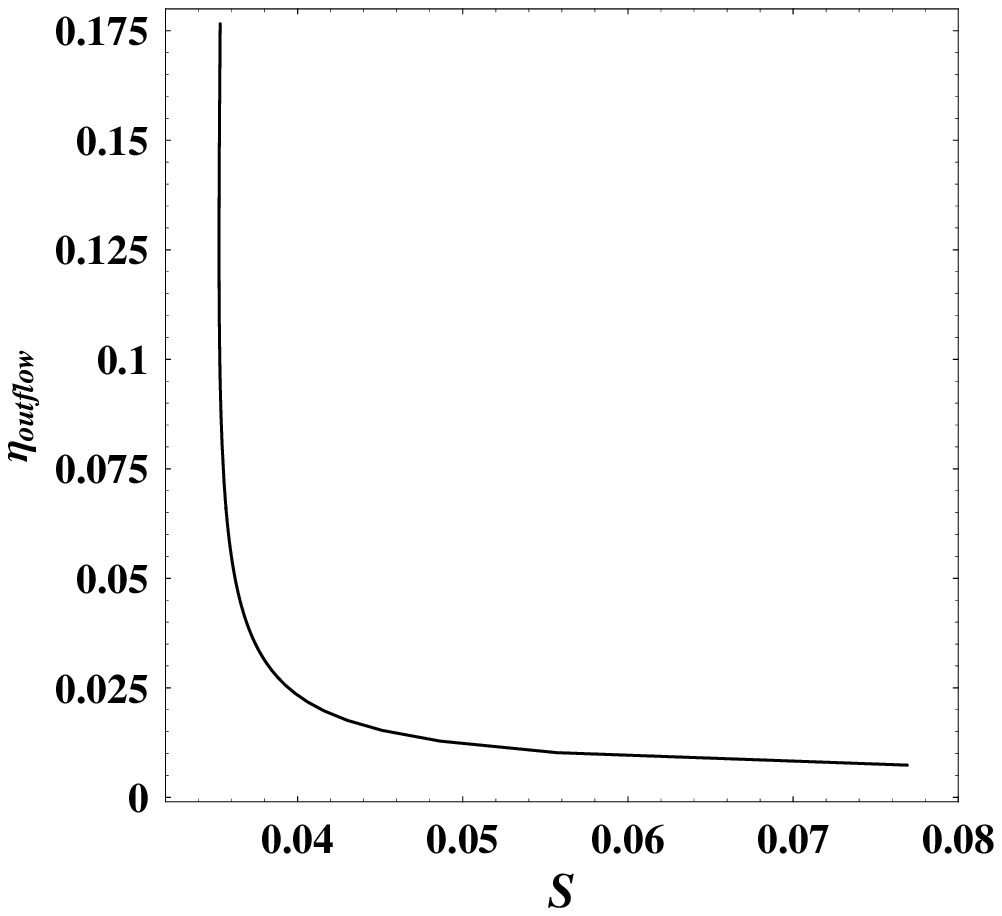} \hfill
\includegraphics[width=5.0cm]{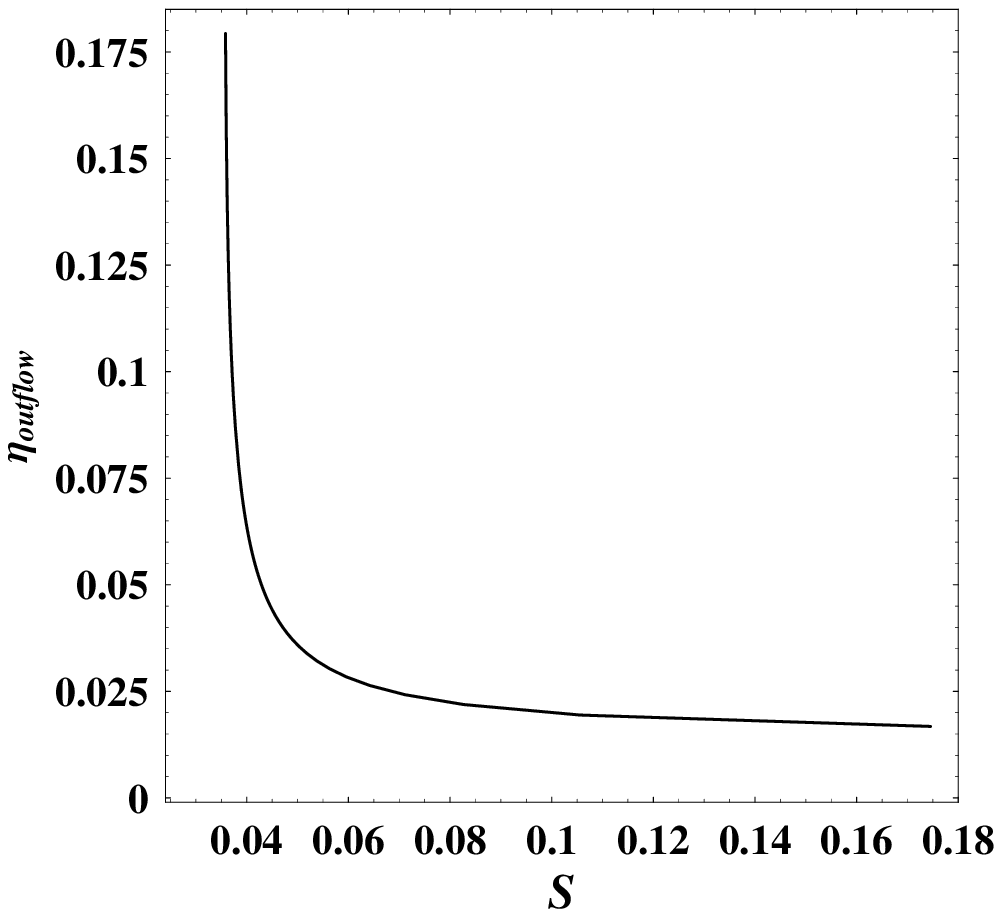} \hfill
\includegraphics[width=5.0cm]{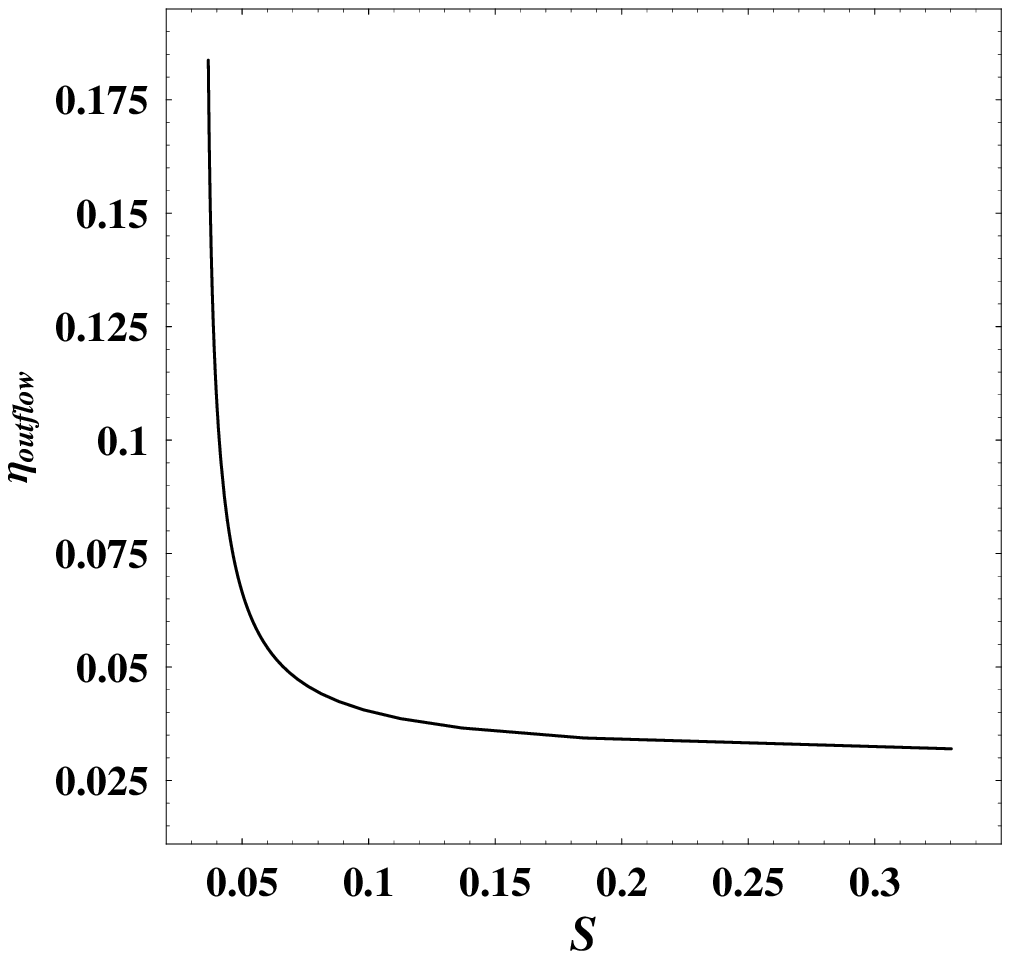}
\centerline{\hspace{0.6cm}(a)\hspace{6cm}(b)\hspace{6cm}(c)} }
\caption{The curves of $\eta _{outflow} $ versus $S$ for different
values of $a_{*}$ and $n$, (a) $a_ * = 0.40$, $n = 7.6$ (b) $a_ * =
0.60$, $n = 6.6$ and (c) $a_* = 0.80$, $n = 5.6$.} \label{fig4}
\end{center}
\end{figure*}

\begin{figure*}
\begin{center}
{\includegraphics[width=5.0cm]{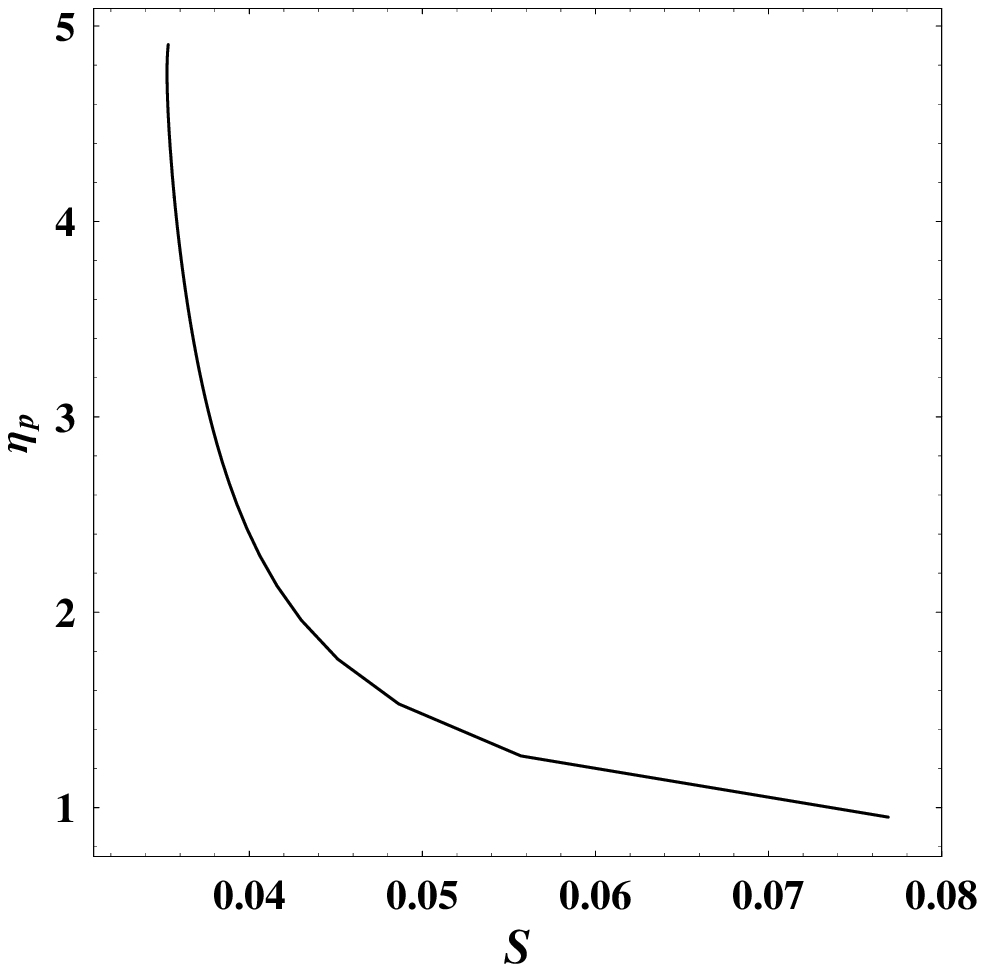} \hfill
\includegraphics[width=5.0cm]{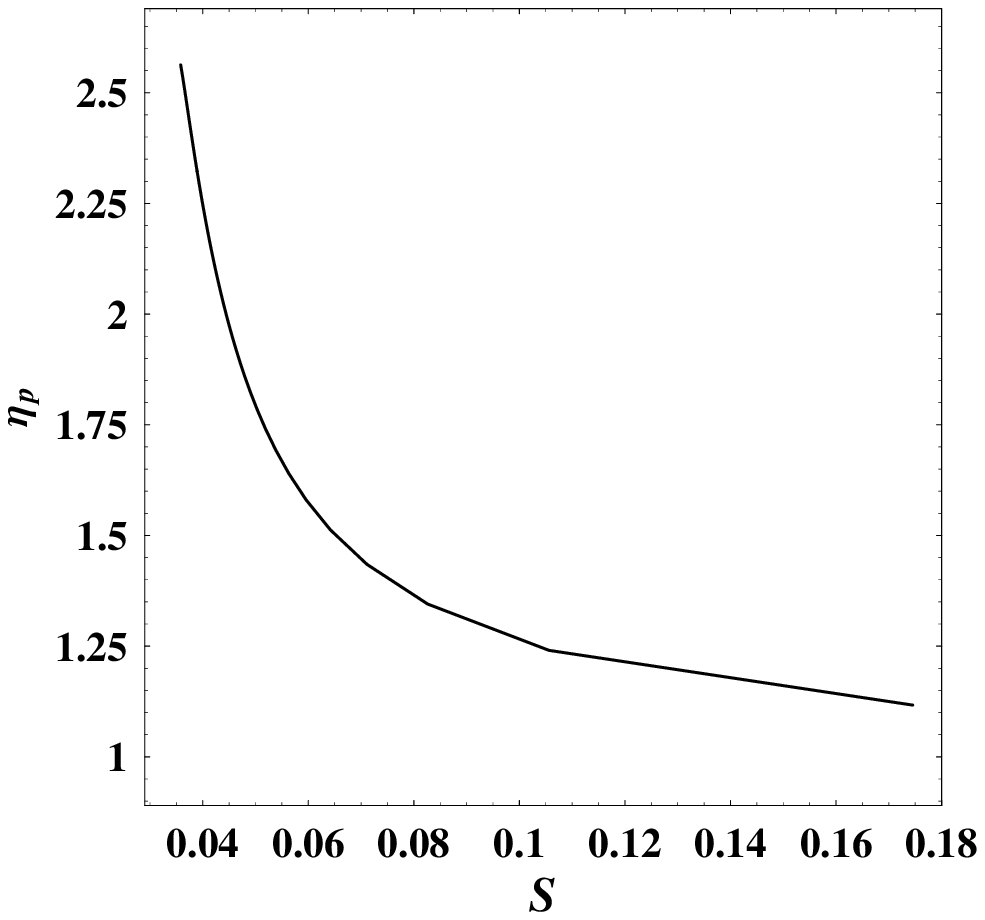} \hfill
\includegraphics[width=5.0cm]{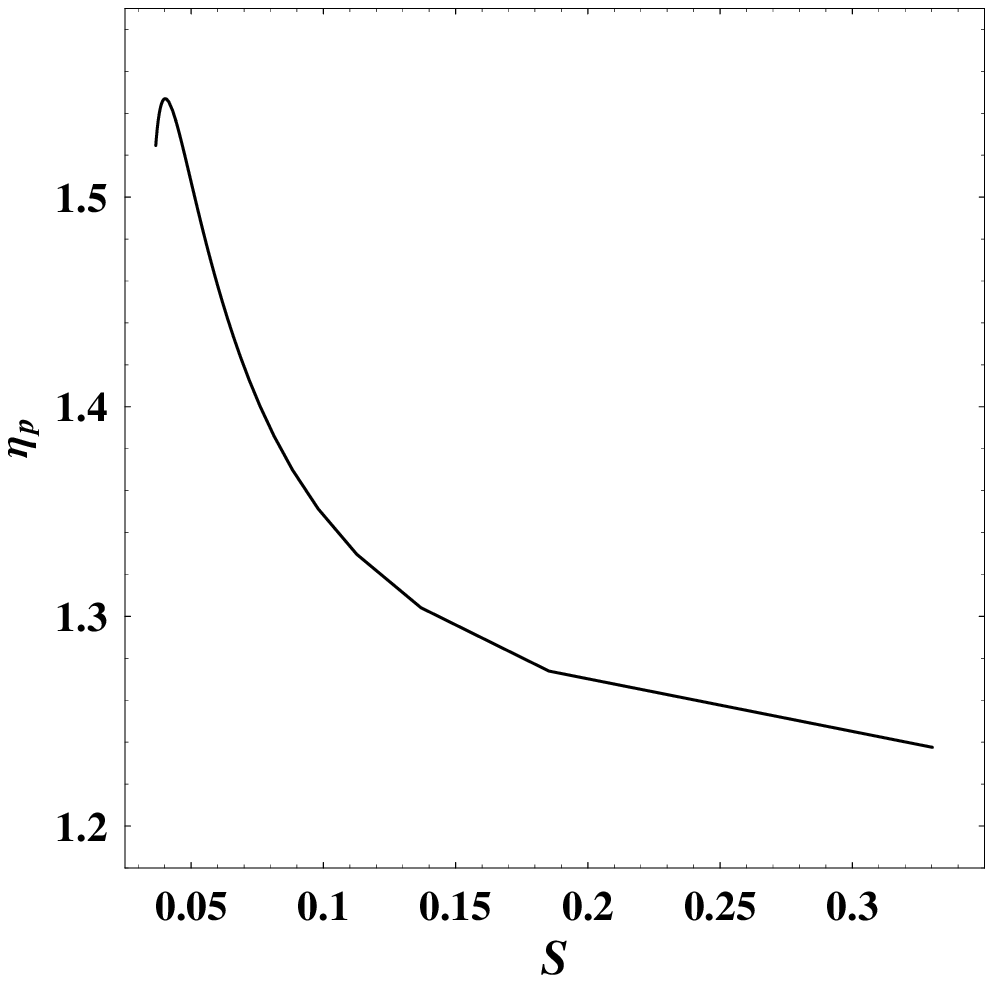}
\centerline{\hspace{0.6cm}(a)\hspace{6cm}(b)\hspace{6cm}(c)} }
\caption{The curves of $\eta _{_{P}} $ versus $S$ for different
values of $a_{*}$ and $n$, (a) $a_ * = 0.40$, $n = 7.6$ (b) $a_ * =
0.60$, $n = 6.6$ and (c) $a_* = 0.80$, $n = 5.6$.} \label{fig5}
\end{center}
\end{figure*}

\begin{figure*}
\begin{center}
{\includegraphics[width=5.0cm]{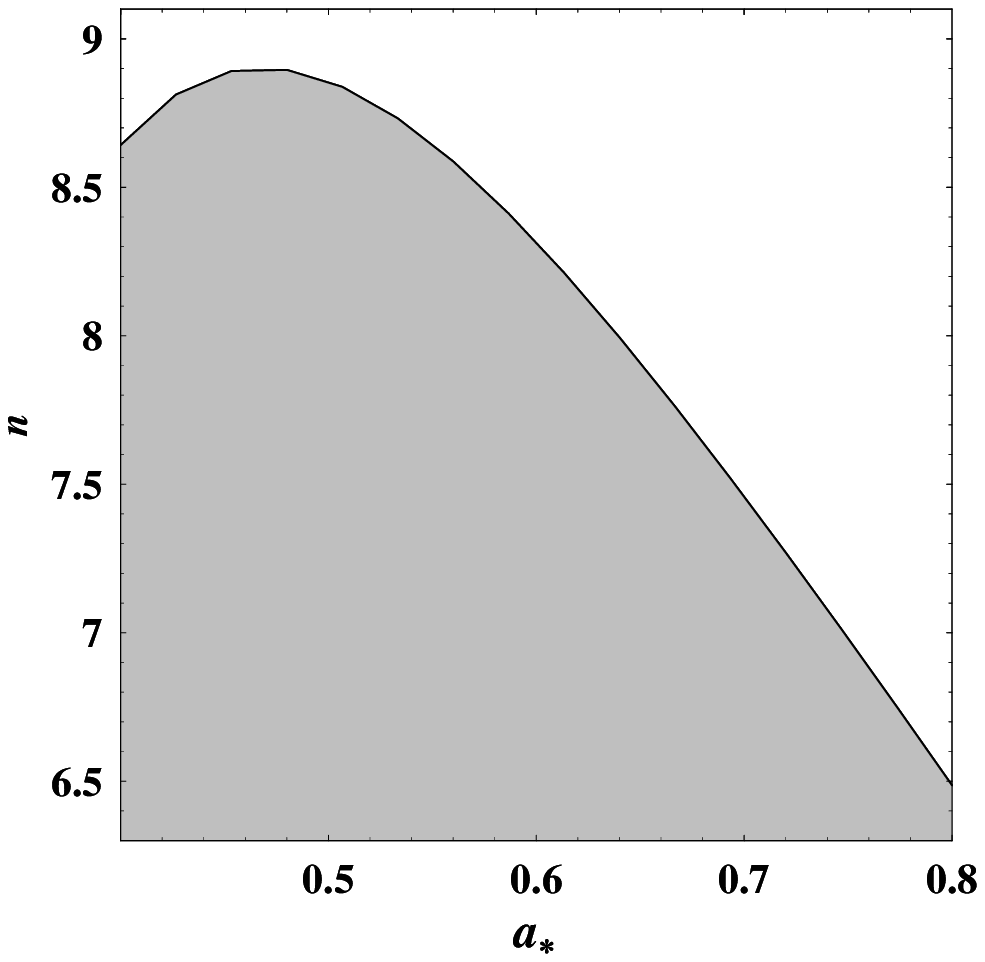} \hfill
\includegraphics[width=5.0cm]{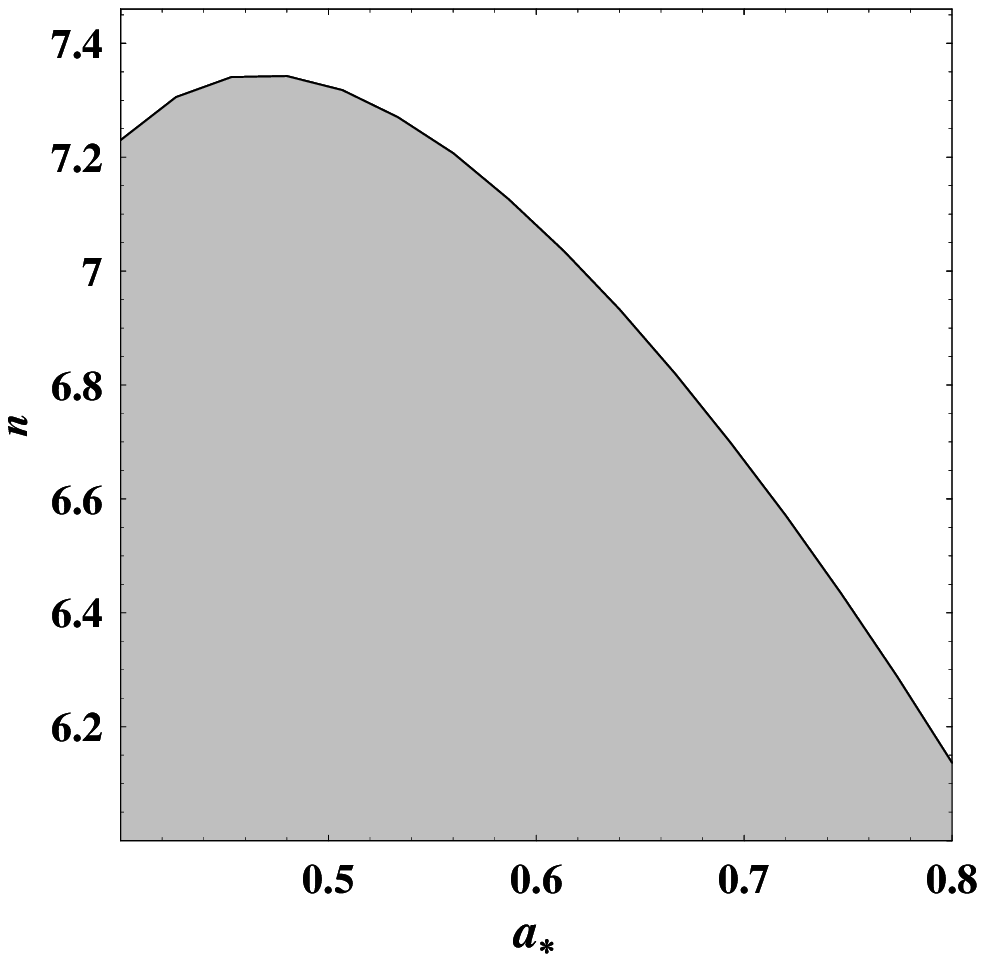} \hfill
\includegraphics[width=5.0cm]{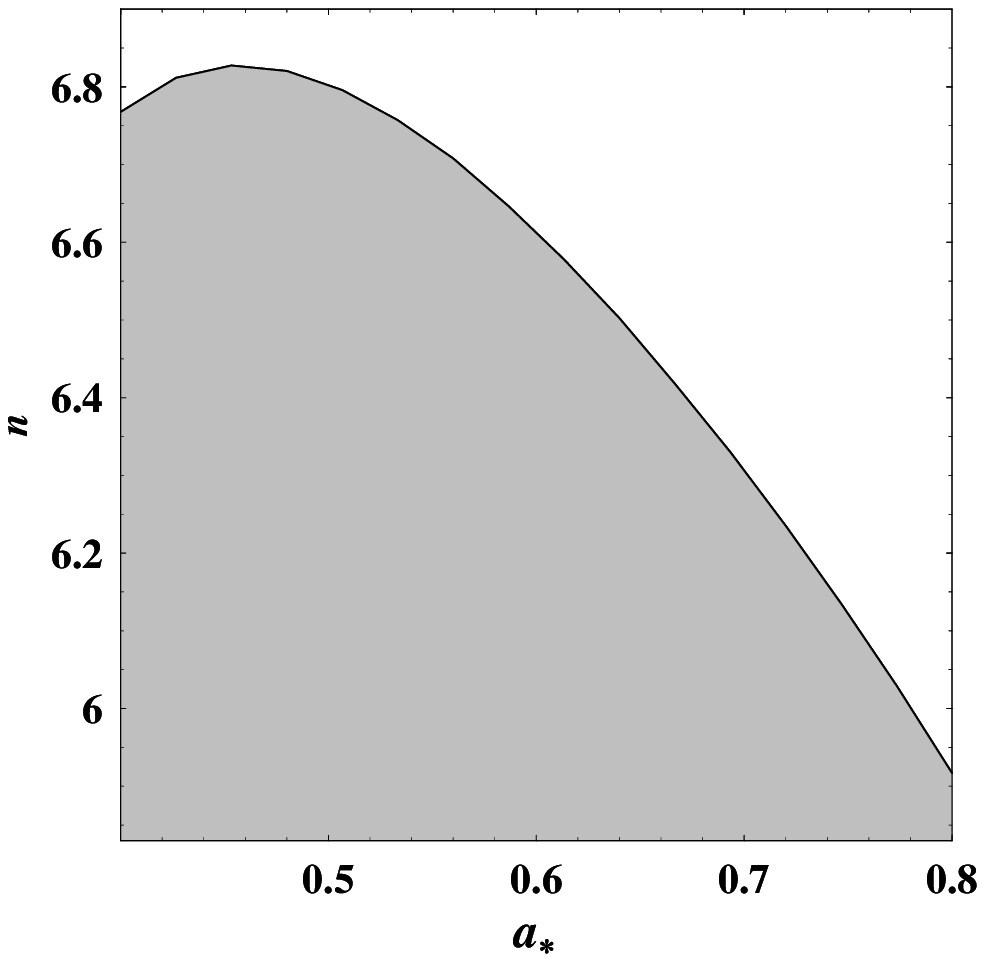}
\centerline{\hspace{0.6cm}(a)\hspace{6cm}(b)\hspace{6cm}(c)} }
\caption{The inequality $\tilde {P}_{BP}>\tilde {P}_{BZ}$ with $\eta
_{_{P}} >1$ holds in the shaded region for (a) $S = 0.2$, (b) $S =
0.4$ and (c) $S = 0.6$.} \label{fig6}
\end{center}
\end{figure*}

\begin{table*}
\caption{The powers and torques in the BZ, MC and BP model with
$\varsigma _{out} = 10^2$}
\begin{tabular}
{|p{48pt}|p{48pt}|p{48pt}|p{48pt}|p{48pt}|p{53pt}|p{53pt}|p{53pt}|}

\hline $a_ * $& $n$ & $h_{c} $& $\tilde {T}_{BZ} $ & $\tilde
{T}_{BP} $ & $\tilde {P}_{BZ} \left( {\times 10^{ - 3}} \right)$&
$\tilde {P}_{BP} \left( {\times 10^{ - 3}} \right) $&$\tilde
{P}_{jet} \left( {\times 10^{ - 3}} \right)$
 \\
\hline 0.40& 7.6& 52.41& 0.077& 12.57& 4.00& 19.73&
23.73 \\
\hline 0.50& 7.1& 41.55& 0.089& 11.42& 5.94& 20.36&
26.30 \\
\hline 0.60& 6.6& 33.73& 0.098& 10.14& 8.18& 20.81&
28.99 \\
\hline 0.70& 6.1& 27.40& 0.107& 8.68& 10.89& 21.18&
32.07 \\
\hline 0.80& 5.6& 21.55& 0.118& 6.97& 14.69& 21.71&
36.40 \\
\hline
\end{tabular}
\label{tab3}
\end{table*}

It is assumed that the accretion rate $\left( {\dot {M}_d }
\right)_{out} $ at $r_{out} $ is related to $\left( {\dot {M}_d }
\right)_{MC} $ by

\begin{equation}
\label{eq36} \left( {\dot {M}_d } \right)_{out} \left( {r^2\omega }
\right)_{out} - \left( {\dot {M}_d } \right)_{MC} \left( {r^2\omega
} \right)_{SIMC} = T_{BP} - \delta T_{MC} ,
\end{equation}

\noindent where $\delta T_{MC} = T_{MC}^{in} - T_{MC}^{out} $ is the
angular momentum transferred from region III to region II, and
$\delta $ is a fraction parameter taken as $\delta = 0.5$ in
calculations. The terms $T_{MC}^{in} $ and $T_{MC}^{out} $ are the
transfer rates of angular momentum at $r_{SIMC} $ and $r_{out} $,
respectively. Substituting equation (15) into equation (40), we have

\begin{equation}
\label{eq371} \alpha _m B_H^2 r_H^2 \left[ {\varsigma _{out}^S
\left( {r^2\omega } \right)_{out} - \left( {r^2\omega } \right)_{MC}
} \right] = T_{BP} - \delta T_{MC} .
\end{equation}

Equation (\ref{eq371}) can be rewritten as

\begin{equation}
\label{eq38} \alpha _m \left( {1 + q} \right)^2\chi _{ms} \varsigma
_{out}^S \left( {\xi _{out}^{1 \mathord{\left/ {\vphantom {1 2}}
\right. \kern-\nulldelimiterspace} 2} - \xi _{SIMC}^{1
\mathord{\left/ {\vphantom {1 2}} \right. \kern-\nulldelimiterspace}
2} } \right) = \tilde {T}_{BP} - \delta \tilde {T}_{MC} .
\end{equation}

Equations (33)---(42) provide a closed set for calculating the
powers and torques in the BZ, MC and BP processes, in which seven
parameters are $a_ * $, $n$, $h_c $, $\alpha _m \quad \delta $, $S$
and $\varsigma _{out} $ are involved. Besides the parameters $\alpha
_m = 0.1$ and $\delta = 0.5$ the rest five parameters are divided
into two types.

(1) The parameters $a_ * $, $n$ and $h_c $ are required to be
greater than some critical values for the coexistence of SIBZ and
SIMC as argued in W06, by which the boundary radii $\xi _{SIMC} $
and $\xi _{SIBZ} $ can be determined.

(2) The parameters $S$ and $\varsigma _{out} $ are involved in the
BP process, and only one of them is independent as shown in Figure 3
below.

Thus we have four independent parameters ($a_ * $, $n$, $h_c $ and
$\varsigma _{out} )$ as the input parameters, and $\xi _{SIMC} $,
$\xi _{SIBZ} $ and $S$ as the derived parameters. Based on the above
analysis we have the values of the derived parameters corresponding
to the input ones as listed in Table 2.

Based on equations (34), (39) and (42) we obtain the curves of the
parameter \textbf{\textit{S}} varying with $\varsigma _{out} $ for
the given values of $a_ * $ and $n$ as shown in Figure 3, and we
find that \textbf{\textit{S}} decreases monotonically with the
increasing $\varsigma _{out} $.

Incorporating equations (16) and (18), we have the ratio of $\dot
{M}_{outflow} $ to $\dot {M}_d $ expressed by

\begin{equation}
\label{eq39} \eta _{outflow} = {\dot {M}_{outflow} } \mathord{\left/
{\vphantom {{\dot {M}_{outflow} } {\left( {\dot {M}_d } \right)_{MC}
}}} \right. \kern-\nulldelimiterspace} {\left( {\dot {M}_d }
\right)_{MC} } = \varsigma _{out}^S - 1,
\end{equation}

\noindent and the curves of $\eta _{outflow} $ versus
\textbf{\textit{S}} for the given values of $a_ * $ and $n$ are
shown in Figure 4.

Inspecting Figure 4, we find that $\eta _{outflow} $ decreases
monotonically with the increasing \textbf{\textit{S}}, and this
result implies that the greater ratio $\varsigma _{out} $
corresponds to the less \textbf{\textit{S }}and the greater outflow
rate.


\section{JET POWER FROM BH ACCRETION DISCS }

As argued above the jets are driven respectively in the Poynting
flux and hydromagnetic regimes by the BZ and BP processes, and the
jet power can be estimated as the sum of the BZ and BP powers,

\begin{equation}
\label{eq40} P_{jet} = P_{BZ} + P_{BP} .
\end{equation}

Thus we have the powers and torques in the BZ-MC-BP model as shown
in Table 3.

The results listed in Table 3 correspond to $\varsigma _{out} =
10^2$, and we find that the BP power is greater than the BZ power,
while the BP torque is about two orders greater than the BZ torque.
Based on equations (33)---(39) we have the curves of $\eta _{_P}
\equiv {P_{BP} } \mathord{\left/ {\vphantom {{P_{BP} } {P_{BZ} }}}
\right. \kern-\nulldelimiterspace} {P_{BZ} }$ versus
\textbf{\textit{S}} for the given values of $a_ * $ and $n$ as shown
in Figure 5.

Inspecting Figure 5, we find that the ratio $\eta _{_P} > 1 $ holds
for a variety of values of the parameters, $a_*$, $n$ and
\textbf{\textit{S}}. Combining Figure 5 with Figure 3, we find that
the BP power can be more stronger than the BZ power for less
\textbf{\textit{S}} and thus for greater $\varsigma _{out}$.

In order to discuss the relative importance of the BP power relative
to the BZ power in a visual way we plot the contour of $\eta _{_P} =
1$ in $a_ * - n$ parameter space for the given values of
\textbf{\textit{S}} as shown in Figure 6, in which the shaded
regions indicate $\eta _{_P} > 1$ for the BP power greater than the
BZ power.

Incorporating Figures 5 and 6, we find that the BP power could be
greater than the BZ power for a wide range of the parameters$a_ * $,
$n$ and \textbf{\textit{S}}. These results are consistent with the
previous works in estimating the BZ power relative to the jet power
from the inner disc (Ghosh {\&} Abramowicz 1997; Livio, Ogilvie {\&}
Pringle 1999).

Cao {\&} Rawlings (2004, hereafter CR04) estimated the jet powers of
a sample of 3CR FR I radio galaxies, and they argued that the BZ
mechanism provides insufficient power to explain the high radio
luminosities of at least a third, and perhaps all, of the sample, if
the accretion discs in these sources are assumed to be advection
dominated accretion flows (ADAFs), or adiabatic inflow-outflow
solution (ADIOS) flows. However, the BP power was not considered in
CR04. From Table 3 we find that $P_{jet} = \left( {2.5\sim 6}
\right)P_{BZ} $, and the jet powers can be amplified significantly
by using equation (44).

\section{DISCUSSION}

In this paper we incorporate three mechanisms into the BZ-MC-BP
model, in which the jet can be powered by both BZ and BP mechanisms,
which belong to the Poynting flux and the hydromagnetic regimes,
respectively. In this model the ratio of the BP power to the BZ
power varies with the concerned parameters, $a_ * $, $n$ and $S$, as
shown in Table 3 and Figure 5.

As pointed out by Meier (2003), there are observational reasons for
believing that the same source may produce jets of rather different
Lorentz factors, either simultaneously or in different accretion
states. The results obtained in this model could interpret the jets
of different Lorentz factors, and are consistent with a spine/sheath
jet structure. The Poynting flux powered by the BZ process
corresponds to the spine with higher Lorentz factor near the axis
and hydromagnetic outflow powered by the BP process corresponds to
the sheath with lower Lorentz factor away from the axis.

Another feature of this model is the role of the MC process. It is
assumed that disc accretion is depressed due to the transfer of the
angular momentum from a rotating BH to the inner disc. It is the MC
effects that give rise to the variation of the accretion rate as
well as the outflow in region II. It has been argued that a very
steep emissivity index required by broad Fe K$\alpha $ lines can be
interpreted by invoking the MC process, which is consistent with the
\textit{XMM-Newton} observation of the nearby bright Seyfert 1
galaxy MCG-6-30-15 (Wilms 2001; Li 2002c; W03).

Recently, Sambruna et al. (2006) discussed the jet/accretion
connection based on Chandra and XMM-Newton observations of three
powerful radio-loud quasars, 1136-135, 1150+497 (\textit{Chandra}),
and 0723+679 (\textit{XMM-Newton}), and they concluded that both jet
and disc emission contribute to the X-ray emission from the three
quasar cores. It was pointed out that the beamed emission
contributes roughly 50{\%} to the total flux in 2-10 keV, while the
disc emission dominates below 2 keV. We expect that the jet
/accretion connection in these radio-loud quasars could be
interpreted based on the BZ-MC-BP model, in which the jet power is
equal to the sum of the BZ and BP powers, and the MC power plus the
disc accretion power can be used to fit the disc luminosity. In
addition, the different Lorentz factor in the jet can be fitted by
adjusting the ratio of the BP power to the BZ power, and the broad
Fe K$\alpha $ lines can be simulated by invoking the MC process as
argued in W03.

Very recently, Wang, Ye {\&} Huang, (2007, hereafter W07) fitted the
twin peak high frequency quasi-periodic oscillations (QPOs)
associated with the jets observed in several sources, in which BH
binary GRO J1655-40 is included. The fitting in W07 is focused on
the association of the 3:2 QPO pairs of the jets based on the
coexistence of the BZ and MC mechanisms, in which the jet is powered
by the BZ process and the MC process is invoked to fit the 3:2 QPO
pairs. It is noted that an X-ray-absorbing wind discovered in an
observation of GRO J1655-40 must be powered by a magnetic process.
Detailed spectral analysis and modeling of the wind shows that it
can only be powered by pressure generated by magnetic viscosity
internal to the disk or magnetocentrifugal forces (Miller et al.
2006). Thus we expect that the fitting given in W07 should be
improved based on the BZ-MC-BP model.

Recently, a lot of works have been done on general relativistic
simulation on BH accretion and outflow (De Villiers, Hawley {\&}
Krolik 2003; Hirose, Krolik, De Villiers {\&} Hawley 2004; De
Villiers, Hawley, Krolik {\&} Hirose 2005; Krolik, Hawley, {\&}
Hirose 2005). McKinney {\&} Narayan (2007) have found that a highly
relativistic, Poynting-dominated funnel jet in the polar regions of
a Kerr BH is associated with a strikingly simple angular-integrated
toroidal current distribution ${dI_\varphi } \mathord{\left/
{\vphantom {{dI_\varphi } {dr}}} \right. \kern-\nulldelimiterspace}
{dr} \propto r^{{ - 5} \mathord{\left/ {\vphantom {{ - 5} 4}}
\right. \kern-\nulldelimiterspace} 4}$, and the polar field is
confined/collimated by the corona. It is interesting to note that
the BZ-MC-BP model is consistent with the simulations in some main
features: (\ref{eq1}) Poynting-dominated jets driven by the BZ
process, and (\ref{eq2}) a large-scale poloidal magnetic field
varying with the disc radius as $B_d^p \propto r_d ^{ - n}$
corresponds to a simple toroidal current with a power-law
distribution in the disc. We shall improve this model based on the
constraints of the numerical simulations as well as the observations
in our future work.

\section*{Acknowledgements}

This work is supported by the National Natural Science Foundation of
China under grants 10573006, 10703002 and 10121503. The anonymous
referee is thanked for his/her helpful comments.

\end{document}